\documentclass[10pt]{article}
\usepackage{graphicx}
\usepackage{amsmath}

\input{tcilatex}

\begin{document}

\title{Einstein's theory of wavefronts versus \\
Einstein's relativity of simultaneity}
\author{Dr. Yves Pierseaux \\
ULB ypiersea@ulb.ac.be}
\maketitle

\begin{abstract}
The relativity of simultaneity implies that the Lorentz transformed (LT)
image of a spherical (circular) wavefront is not a spherical (circular)
wavefront (Einstein 1905) but an ellipsoidal (elliptical) wavefront \cite{0}%
. We show first that the relativity of simultaneity leads to the consequence
that the image of a Lorentz transformed plane wavefront is a tangent plane
to an ellipsoid and not a tangent plane to a sphere (Einstein 1905). We then
deduce a longitudinal component of the tangent vector to Poincar\'{e}'s
ellipse which is directly connected to the relativity of simultaneity. We
suggest finally that this violation of simultaneity is related to Einstein's
implicit choice of the (non relativistic) transverse gauge.
\end{abstract}

\section{Relativity of simultaneity (LT) and Elliptical wavefront}

Let us consider two inertial systems K and K' in uniform translation
relative to one another. A source of light is at rest O in K. What is the
image by a LT in K' in $t_{0}$, to a circular wavefront in K, emitted at $%
t^{\prime }=t=0$ by this source S? Let us write the LT with perfectly
spacetime symmetry (in x and t) \cite{4}:

\begin{equation}
x^{\prime }=\gamma (x+\beta t)\qquad \qquad \qquad y^{\prime }=y\qquad
\qquad t^{\prime }=\gamma (t+\beta x)
\end{equation}

\bigskip Let us consider the relativistic invariant: 
\begin{equation}
x^{2}+y^{2}=r_{0}^{2}=t_{0}^{2}\qquad \qquad x^{\prime 2}+y^{^{\prime
}2}=r^{\prime 2}=t^{\prime 2}
\end{equation}

The object time $t=t_{0}$ is fixed in K (circular wavefront in K) but\ the
image time \textit{t'} is not fixed (by the LT) \textit{in} K'. We obtain
using (1) $t^{\prime }=\gamma ^{-1}t_{0}+\beta x^{\prime }$: 
\begin{equation}
x^{\prime 2}+y^{^{\prime }2}=(\gamma ^{-1}t_{0}+\beta x^{\prime })^{2}\qquad 
\text{or}\qquad (\gamma ^{-1}x^{\prime }-\beta t_{0})^{2}+y^{\prime
2}=t_{0}^{2}
\end{equation}
which is the Cartesian equation of an elongated ellipse\ in K' with the
observer O' at the focus, \textbf{figure 1}):

The physical meaning of this elliptical spacetime wavefront is the \textbf{%
relativity of simultaneity}: two simultaneous events (\textit{different
abscissa}) in K are not simultaneous in K' \cite{2} : 
\begin{equation}
\Delta t=0\qquad \rightarrow \qquad \Delta t^{\prime }=\frac{t^{\prime
+}-t^{\prime -}}{2}\neq 0
\end{equation}

$\Delta t^{\prime }$ is \textit{''the simultaneity gap''} between two
opposite events on the front (\textbf{figure 1}). If according to Einstein 
\cite[paragraphe 3]{1}, the object (fixed time t) and the image (fixed time
t'), are both spherical (circular)\footnote{%
Einstein writes :'' At the time $t=\tau =0$, when the origin of the two
coordinates (K and k) is common to the two systems, let a \textbf{spherical
wave }be emitted therefrom, and be propagated -with the velocity c in system
K. If x, y, z be a point just attained by this wave, then $%
x^{2}+y^{2}+z^{2}=c^{2}t^{2}.$ Transforming this equation with our equations
of transformation (see Einstein's LT, 29), we obtain after a simple
calculation $\xi ^{2}+\eta ^{2}+\zeta ^{2}=c^{2}\tau ^{2}.$ The wave under
consideration is therefore no less a \textbf{spherical wave }with velocity
of propagation c when viewed in the moving system k.''\cite{1}}, then\ two
simultaneous events in K must be always also simultaneous in K': 
\begin{equation}
(\Delta t^{\prime }=0)_{\text{Einstein}}
\end{equation}
In polar coordinates ($x^{\prime }=r^{\prime }\cos \theta ^{\prime },$ $%
y^{\prime }=r^{\prime }\sin \theta ^{\prime },$with $\theta ^{\prime }$ as
polar angle and F as pole) the equation of the elongated ellipse is :

\begin{equation}
r^{\prime }=\frac{r_{0}}{\gamma (1-\beta \cos \theta ^{\prime })}
\end{equation}

\bigskip With relativistic transformation of angle \cite{0} 
\begin{equation}
\cos \theta ^{\prime }=\frac{\cos \theta +\beta }{1+\beta \cos \theta }
\end{equation}

We rediscover the LT for any point of the wavefront $(r,\theta $ or $%
t,\theta )$ :

\begin{equation}
r^{\prime }=\gamma r_{0}(1+\beta \cos \theta )\qquad \qquad t^{\prime
}=\gamma t_{0}(1+\beta \cos \theta )
\end{equation}

The velocity of light (''one way'') is identical in all directions within
both systems because we have by construction \textbf{figure 1}):$\frac{r_{0}%
}{t_{0}}=\frac{r^{\prime +}}{t^{\prime +}}=\frac{r^{\prime -}}{t^{\prime -}}%
=c=1.$ What is a wavefront in a relativistic sense or in other words a
''spacetime'' wavefront? The invariant interval between the event
''emission'' $(t=t^{\prime }=0)$ and any event on the wavefront $(t=t_{0},$ $%
r=r_{0}$ in K and $\ \ t^{\prime },$ $r^{\prime }$ in K'$)$ is null within
both systems. So Minkowski's null spacetime 4-vector is, from Poincar\'{e}'s
point of view\footnote{%
Poincar\'{e} first introduced \cite{5} the ellipsoidal image of a spherical
wavefront. However Poincar\'{e}'s historical ellipse (source at the focus)
is not equation (6). It is given by the inverse LT by changing (in 6) the
sign of the velocity and by inverting prime and non primed.}, a wavefront
4-vector\textit{. The elliptical image by\ a LT is deduced from a }\textbf{%
non-transversal }$\mathit{\ }t^{\prime }$\textit{\ \textbf{section }in
Minkowski's cone. }

According to Einstein, the image of a space wavefront (t is fixed) is a
space wavefront (t' is fixed). We will see now that is exactly the same for
Einstein's plane wavefront \cite[paragraphe 7]{1}.

\section{Relativity of simultaneity (LT) and the tangent plane wavefront to
the ellipsoid}

Until now we considered only one source $S_{O}$ which emits at $t=t^{\prime
}=0$ a spherical wave at fixed time $t=t_{0}$ (figure 1)- Let us now
consider a second source $S_{\infty }$ at infinity at rest in K in the
direction $\theta $ (\textbf{figure 2}) which emits a plane wavefront.
Suppose that the considered plane wavefront (here a ''wave straight line''
at two space dimensions) be in O at the time $t=t^{\prime }=0$ (when $S_{O}$
emits a spherical wave). It will be \textbf{tangent} in $t=t_{0}$ to the
circular wavefront, emitted by $S_{1}$ (normalized $t_{0}=r_{0}=1$ \textbf{%
figure 2}). We note that the simultaneous events $P_{1}TP_{2}$ in K are no
longer simultaneous\textit{\ }($P_{1}^{\prime }T^{\prime }P_{2}^{\prime }$)
in K'. We deduce respectively (1) for the object front (''wave straight
line'' with angular coefficient $a=-\cot $ $\theta $) and the image front
thefollowing relations:

\begin{equation}
x\cos \theta +y\sin \theta =t_{0}\qquad \qquad y^{\prime }\sin \theta
^{\prime }+x^{\prime }\cos \theta ^{\prime }=t^{\prime }
\end{equation}

And exactly as in the circular wavefronts case (2), there are two
possibilities:

\textbf{If }$\mathbf{t}^{\prime }$\textbf{\ \ is not fixed (Poincar\'{e}, }%
spacetime image wavefront\textbf{),} we have by a LT, $t^{\prime }=\gamma
(t+\beta x)=\gamma ^{-1}t+\beta x^{\prime }$, and therefore the primed
equation (9) is the equation of the tangent to the ellipse at the point $%
T^{\prime }$ (on which $P_{1}^{\prime }$ and $P_{2}^{\prime }$ are situated$%
) $ :

\begin{equation}
y^{\prime }\sin \theta ^{\prime }+x^{\prime }(\cos \theta ^{\prime }-\beta
)=\gamma ^{-1}t
\end{equation}

The angular coefficient of Poincar\'{e}'s wavefront:

\QTP{Body Math}
\begin{equation}
a_{poincar\acute{e}}^{\prime }=tg\alpha ^{\prime }=\frac{\beta -\cos \theta
^{\prime }}{\sin \theta ^{\prime }}
\end{equation}

\textit{\ If \ }$\mathbf{t}^{\prime }$\textbf{\ }\textit{\textbf{is fixed} 
\textbf{(Einstein, }}space image wavefront\footnote{%
Einstein defined clearly the image of the wavefront in paragraph 7 of his
1905 paper: ''If we call the angle $\theta ^{\prime }$the angle between 
\textbf{the wave-normal (direction of the ray)} and the direction of
movement''.}\textit{\textbf{) }the primed equation} (9) is the equation of
the tangent to the circle (with centre O' and radius $r_{T}^{\prime })$. The
angular coefficient of Einstein's wavefront is: 
\begin{equation}
a_{einstein}^{\prime }=-\cot \theta ^{\prime }
\end{equation}

Both formulas are the same for $\theta ^{\prime }=0.$ \textbf{Einstein's} 
\textbf{double transversality} (or \textbf{Einstein's double simultaneity } $%
t_{fixed}\neq $ $t_{fixed}^{\prime }$ ) involves for the image front:\textit{%
\ } 
\begin{equation}
a^{\prime }=-\cot \text{ }\theta ^{\prime }\qquad \Leftrightarrow \qquad
(\Delta t^{\prime })_{front}=0
\end{equation}

This is consistent with (5): Einstein's plane image wavefront is tangent to
Einstein's spherical image wavefront. The phase $\Psi $ of a sinusoidal
monochromatic plane wave ($\mathbf{A}$ is the amplitude) $\mathbf{A}=\mathbf{%
A}_{0}\sin \Psi $ ($\mathbf{A}^{\prime }=\mathbf{A}_{0}^{\prime }\sin \Psi
^{\prime })$ is defined by the 3-scalar product $\mathbf{k.r}$ ($\mathbf{k}%
^{\prime }.\mathbf{r}^{\prime }\mathbf{)}$ with the frequency $\nu =\frac{%
\omega }{2\pi }$, the wave vector $\mathbf{k}=\frac{2\pi }{\lambda ^{\prime }%
}\mathbf{1}_{n}$ with $\mathbf{1}_{n}$ the unit vector normal to the front ($%
\mathbf{k}^{\prime }=\frac{2\pi }{\lambda ^{\prime }}\mathbf{1}_{n^{\prime
}},\nu ^{\prime }=\frac{\omega ^{\prime }}{2\pi }):$ 
\begin{equation}
\omega t-\mathbf{k.r=}\text{ }\Psi =\omega ^{\prime }t^{\prime }-\mathbf{k}%
^{\prime }\mathbf{.r}^{\prime }=\Psi ^{\prime }\qquad \Leftrightarrow \qquad 
\mathbf{A}.\mathbf{k=A}^{\prime }.\mathbf{k}^{\prime }=0=Ak\cos \phi
=A^{\prime }k^{\prime }\cos \phi ^{\prime }
\end{equation}

This is a \textbf{Galilean invariant} (t is fixed on the object front and t'
is fixed on the image front) in the sense that the angle (\textbf{figure 2}) 
$\phi =\phi ^{\prime }=90%
{{}^\circ}%
$ is not altered by a Galilean transformation (GT).

\begin{equation}
\mathbf{k.r}\ \qquad \underrightarrow{GT}\ \qquad \mathbf{k}^{\prime }.%
\mathbf{r}^{\prime }\mathbf{\ \ \ \ \ \ \ \ }\text{or\qquad }\mathbf{A}\perp 
\mathbf{k}\ \qquad \underrightarrow{GT}\ \qquad \mathbf{A}^{\prime }\perp 
\mathbf{k}^{\prime })
\end{equation}
But LT changes (11) this angle $\phi =90%
{{}^\circ}%
$ into $\phi ^{\prime }$ in the following way (in \textbf{figure 2, }we note 
$\pi -\alpha ^{\prime }$): 
\begin{equation}
\tan \phi ^{\prime }=\tan (\alpha ^{\prime }-\theta ^{\prime })=\frac{\beta
\cos \theta ^{\prime }-1}{\beta \sin \theta ^{\prime }}
\end{equation}

The right angle $\phi =\phi ^{\prime }=90%
{{}^\circ}%
$ is conserved \textit{if and only if }the propagation is purely
longitudinal ($\sin \theta ^{\prime }=0)$. Einstein considers 
\cite[paragraphe 7]{1} that the Galilean invariant $t_{fixed}=$ $%
t_{fixed}^{\prime }$ (14) is \textit{by definition} a Lorentz invariant $%
t_{fixed}\neq $ $t_{fixed}^{\prime }$.\footnote{%
Everything happens as if $\beta =0"$, in other words exactly as in
Einstein's synchronisation ''in spherical waves'' \cite{3}.}

\section{Poincar\'{e}'s longitudinal component of a vector tangent to the
ellipse}

Let us now consider a vector $\mathbf{T=}\overrightarrow{TP}$ on the tangent
to the circle object. What is the image $\mathbf{T}^{\prime }$ by LT on the
tangent to the ellipse of $\mathbf{T}$?

\begin{equation}
x_{T^{\prime }}^{\prime }=\gamma (x_{T}+\beta t)\qquad y_{T}^{\prime
}=y_{T}\qquad t_{T^{\prime }}^{\prime }=\gamma (t+\beta x_{T})\qquad \qquad
x_{P^{\prime }}^{\prime }=\gamma (x_{P}+\beta t)\qquad y_{P}^{\prime
}=y_{P}\qquad t_{P^{\prime }}=\gamma (t+\beta x_{P})
\end{equation}

We deduced the components ($T_{x^{\prime }}^{\prime }$ $,T_{y^{\prime
}}^{\prime })$ of the vector $\mathbf{T}^{\prime }\mathbf{=}\overrightarrow{%
T^{\prime }P^{\prime }}$ on the tangent to the ellipse (\textbf{figure 3})

\begin{equation}
T_{x^{\prime }}^{\prime }=\Delta x^{\prime }=\gamma \Delta x\qquad \qquad
\qquad T_{y^{\prime }}^{\prime }=\Delta y^{\prime }=\Delta y\qquad \qquad
\Delta t^{\prime }=\beta \gamma \Delta x
\end{equation}

where $\Delta t^{\prime }$ is ''simultaneity gap''.

Let us project the vectors $\mathbf{T}$\ and $\mathbf{T}^{\prime }$ on the
direction of propagation and on the perpendicular to that direction in both
systems K and K' (the origin is respectively in T and in T'). We deduce
immediately from geometrical properties of the ellipse:

\begin{equation}
T=T_{\perp }=T_{\perp }^{\prime }\qquad \qquad \qquad \qquad T_{\parallel
}^{\prime }=\gamma \beta T_{x}=\Delta t^{\prime }
\end{equation}

\textbf{Poincar\'{e}'s theorem}: ''\textit{The simultaneity gap only depends
on the longitudinal component }$T_{\parallel }^{\prime }$\textit{\ of the
vector }$\mathbf{T}$\textbf{'}\textit{\ of the tangent to the ellipse. }%
Einstein cancels Poincar\'{e}'s longitudinal component: 
\begin{equation}
(T_{\parallel }^{\prime }=\Delta t^{\prime }=0)_{einstein}
\end{equation}

Let us show now that we can associate to the spatial vector $T$ and the\ $%
\Delta t$ a \ 4-vector structure $(T_{x},$ $T_{y}\ ,\ 0,$ $\Delta t=0)$ $%
\rightarrow $\ $(T_{x}^{\prime },$ $T_{y}^{\prime },$ $0,$ $\Delta t^{\prime
})$ with invariant norm:

\begin{equation}
\left\| (T_{x},\qquad T_{y}\ ,\ \ \ \ \ 0,\qquad 0)\right\| =T^{2}=T_{\perp
}^{2}
\end{equation}
$\qquad $%
\begin{equation}
\left\| (T_{x}^{\prime },\qquad T_{y}^{\prime },\ \ \ \ \ \ \ 0,\qquad
\Delta t^{\prime })\right\| =T^{\prime 2}-\Delta t^{\prime 2}=T_{\perp
}^{^{\prime }2}+T_{\parallel }^{\prime 2}-\Delta t^{\prime 2}=T_{\perp
}^{\prime 2}
\end{equation}
where the fourth component $\Delta t^{\prime }$ is cancelled in Einstein's
theory of wavefront.

\section{Conclusion}

\bigskip Einstein's theory of (spherical and plane) wavefronts is not
compatible with Einstein's relativity of simultaneity. Einstein's wavefronts
are ''rigid'' in a prerelativistic meaning: a set of simultaneous events is
transformed into a set of simultaneous events. But this unexpected internal
contradiction has possibly\textit{\ no physical consequence}. Indeed we
showed that only the electromagnetic potential (which is 4-vector \textbf{\ }%
written for the first time by Poincar\'{e} in 1905\textbf{),} and not the
electromagnetic field, is sensitive to this violation of relativity of
simultaneity. We showed \cite{6}\cite{7} that Einstein's theory of
wavefronts is based on the\textit{\ transverse} gauge (completed Coulomb
gauge) whilst Poincar\'{e}'s theory of spacetime wavefronts is based upon
the Lorenz gauge. The implicit Einstein's choice\cite[paragraph 7]{1} of the
completed Coulomb gauge (transverse gauge) explains the reason why the
relativity of simultaneity is violated.

\end{document}